\documentclass[aps,prd,preprint,superscriptaddress,tightenlines,nofootinbib]{revtex4}

\usepackage{graphicx}
\usepackage{dcolumn}
\usepackage{bm}

\begin{document}

\preprint{CLNS 05/1907}       
\preprint{CLEO 05-2}         

\title{Limits on Neutral $D$ Mixing in Semileptonic Decays}


\author{C.~Cawlfield}
\author{B.~I.~Eisenstein}
\author{G.~D.~Gollin}
\author{I.~Karliner}
\author{D.~Kim}
\author{N.~Lowrey}
\author{P.~Naik}
\author{C.~Sedlack}
\author{M.~Selen}
\author{J.~Williams}
\author{J.~Wiss}
\affiliation{University of Illinois, Urbana-Champaign, Illinois 61801}
% ...
\collaboration{CLEO Collaboration} 
\noaffiliation

\date{February 5, 2005}

\begin{abstract} 
Using the CLEO II.V detector observing $e^+e^-$ collisions
at around 10.6~GeV we search for neutral $D$ mixing in semileptonic
$D^0$ decays tagged in charged $D^\ast$ decays.
Combining the results from the $Ke\nu$ and
${K^\ast}e\nu$ channels we find that the rate for
$D$ mixing is less than 0.0078 at 90\% C.L. 
\end{abstract}

\maketitle

The study of mixing
in the $K^0$ and $B_d^0$ sectors has provided a wealth of information
to guide the form and content of the Standard Model.  In the framework of the 
Standard Model, mixing in the charm meson sector is predicted to 
be small~\cite{harryandalexey}, making this an excellent place to
search for non-Standard Model effects.

A $D^0$ can evolve into a $\overline{D^0}$ through well known, on-shell 
intermediate states, or through off-shell intermediate states such as
those that might be present due to new physics.  We denote the
amplitude through the former (latter) states by $-iy$ $(x)$, in
units of $\Gamma_{D^0} / 2$~\cite{Mixing}.  
The Standard Model contributions to
$x$ are suppressed to $|x| \approx \tan^2 \theta_C \approx 5\%$ and the
Glashow-Illiopolous-Maiani~\cite{GIM} cancellation could further
suppress $|x|$ down to $10^{-6} - 10^{-2}$.  Many non-Standard Model processes
could lead to $|x| > 1\%$.  Signatures of new physics include $|x| \gg |y|$ and
{\em CP} violating interference between $x$ and $y$ or between $x$ and a 
direct decay amplitude.

	
	Observation of $D$ mixing in hadronic decay channels is complicated
by doubly Cabibbo suppressed decays, where both the decay of the original
charm quark and subsequent charged $W$ decay proceed in Cabibbo suppressed
modes.  Such a decay mimics mixing of a $D^0$ to a $\overline{D^0}$ followed
by the dominant Cabibbo allowed channel decay.  In semileptonic decays no such double
suppression is allowed as the $W$ decays to a charged lepton and neutrino, and the 
charge of the lepton tags whether a $c$, producing
a positive lepton, or ${\bar c}$, producing a negative lepton, has decayed.  When the
production flavor of the $D$ is tagged in charged $D^\ast$ decay or some other way,
a wrong sign lepton produced in a subsequent semileptonic decay unambiguously signals
$D$ mixing.  Other mechanisms that produce leptons of the opposite sign in $D^0$ or
$\overline{D^0}$ decay are highly suppressed.
The integrated mixing rate normalized to the total decay rate
is equal to $\frac{1}{2}(x^2 + y^2)$, and is called $R_M$.
This is at once good and bad news.  The observation
of any wrong sign semileptonic $D$ decay is an unambiguous signal of $D$ mixing.  
With both $x$ and $y$ small ($< {\cal O}(0.01)$) the rate of wrong sign semileptonic $D$ decays
will be very small ($< {\cal O}(0.0001)$), and an observation
of $D$ mixing in semileptonic decays would not give insight
on the relative sizes of $x$ and $y$.  In contrast hadronic decays
have wrong sign contributions from both doubly Cabibbo suppressed decays
and mixing.  The two channels interfere resulting in a term that depends
on a linear combination of $x$ and $y$, and the proper time dependence of wrong sign final states
can be used to measure $x$ and $y$ separately.

The proper decay time dependence of semileptonic mixed final states
in units of the
mean $D^0$ lifetime, $t_{D^0} = (410.3 \pm 1.5) {\rm fs}$~\cite{PDG},
is $r(t) \equiv ({x}^2 + {y}^2 ) t^2  e^{-t}$.
Thus mixed semileptonic decays should populate larger proper times with
an average decay time of three.  The dominant direct decays
are distributed as $e^{-t}$ with an average
decay time of one.  This difference is taken advantage of to increase the
sensitivity to $D$ mixing.

	To date no one has observed evidence for $D$ mixing in any channel.
The best limit on $R_M$ comes from the FOCUS collaboration~\cite{FOCUS}
where they find $R_M < 0.0010$ at 90\% C.L.  Belle~\cite{Belle}
reports a similar limit, $R_M < 0.0014$ at 90\% C.L. and 
BABAR~\cite{BaBar} a higher limit of $R_M < 0.0042$ at 90\% C.L.
Other limits on $D$ mixing are summarized in~\cite{PDG2}.
Our experimental situation is very similar to Belle and BABAR, but
our data sample is more than an order of magnitude smaller.
Thus we expect not to be sensitive to any $D$ mixing signal
and set limits higher than above.

 	Our data were collected using the CLEO II.V upgrade~\cite{ii.v} of the
CLEO II detector~\cite{cleoii} between February 1996 and February 1999 
at the Cornell Electron Storage Ring (CESR).  
The data correspond to $9.0$ fb$^{-1}$ of $e^+ e^-$ collisions near $\sqrt{s} 
\approx 10.6$ GeV.  The detector consisted of cylindrical tracking chambers and
an electro-magnetic calorimeter immersed in a 1.5 Tesla axial
magnetic field, surrounded by muon chambers.  The reconstruction of
displaced vertices from charm decays is made possible by the addition
of a silicon vertex detector (SVX) in CLEO II.V.  
The charged particle trajectories are fit using a Kalman filter 
technique~\cite{kalman} that takes into account energy loss as the particles pass through 
the material of the beam pipe and detector.
Specific ionization for charged particle identification is measured in the main drift chamber with
a resolution of about 7\%.  Electrons above 500 MeV/c momentum
are also identified by matching of track
momentum with calorimeter energy deposition and requiring that the
calorimeter shower has the characteristics expected of an electro-magnetic
rather than a hadronic shower.  Hadrons are misidentified as electrons
at roughly the 0.1\% level from studies of known hadron samples in
the data.  Muons are not used as the CLEO muon identification system
cannot cleanly identify muons below about 1.5~GeV/c momentum and the small
sample of clean muons are not
useful to study semileptonic $D$ decays at our beam energy.

To study backgrounds and relative selection efficiencies
we use a GEANT \cite{GEANT} based detector
simulation of our data.  We use simulations of $e^+e^- \to q \bar q$
with the quarks fragmenting and particles decaying
generically guided by previous measurements.  The data of this generic simulation
corresponds to roughly ten times the luminosity collected by the detector.
We also use simulations of $e^+e^- \to \tau^+\tau^-$ and $e^+e^- \to B \overline{B}$
to study small contributions to the background.
For these studies the simulated
samples are reconstructed and selected using the same methods
as the data sample as described below.


To select hadronic events and ensure that the event production
point is well known there must be at least five well reconstructed tracks
consistent with coming from the interaction region.  The tracks must carry
more than 15\% of the collision energy.  This selection
is nearly 100\% efficient, removes events that would not pass subsequent
reconstruction requirements, and ensures a well measured $D^0$ flight distance.
All tracks used in the reconstruction of the
decay chain except the pion from the charged $D^\ast$ decay are required to
hit at least two of the three SVX planes in projections both transverse
and parallel to the beam direction.


	The $D^0$s are tagged at production in the decay of the
charged $D^\ast$ to a charged pion and a $D^0$.  A $\pi^+$ indicates
that a $D^0$ has been produced and a $\pi^-$, a $\overline{D^0}$.
Subsequent direct semileptonic decays of the $D^0$ or $\overline{D^0}$
produce a charged lepton of the same sign
as the pion from the $D^\ast$ decay, called right sign
(RS) combinations,
while semileptonic $D$ decay after mixing produces a charged lepton
of the opposite sign, called wrong sign (WS) combinations.
Since little energy is released
in the $D^\ast$ decay the pion is limited to a momentum of 400 MeV/c
at our beam energy, and is called the soft pion.
We choose combinations such that the momentum of
the $D^0$ is larger than 2.0 GeV/c and thus
are dominated by $D^0$s produced in $e^+e^- \to c \bar c$ with
only a small contribution produced by $B$ decays.
The analyses find RS combinations maximizing
the signal to noise for the decay chains 
$D^\ast \to \pi_{\rm soft} D^0 \to K^{(\ast)} e \nu$.
The same selections are then used to find WS combinations,
and the ratio of WS to RS combinations after accounting
for background is a measure of $R_M$.
The analysis of the $D^0 \to K e \nu$ channel is
described in full detail in~\cite{sedlack} while
the $D^0 \to K^\ast e \nu$ channel is similarly
described in~\cite{mcgee}. 

\begin{figure}
\includegraphics[height=3.0in]{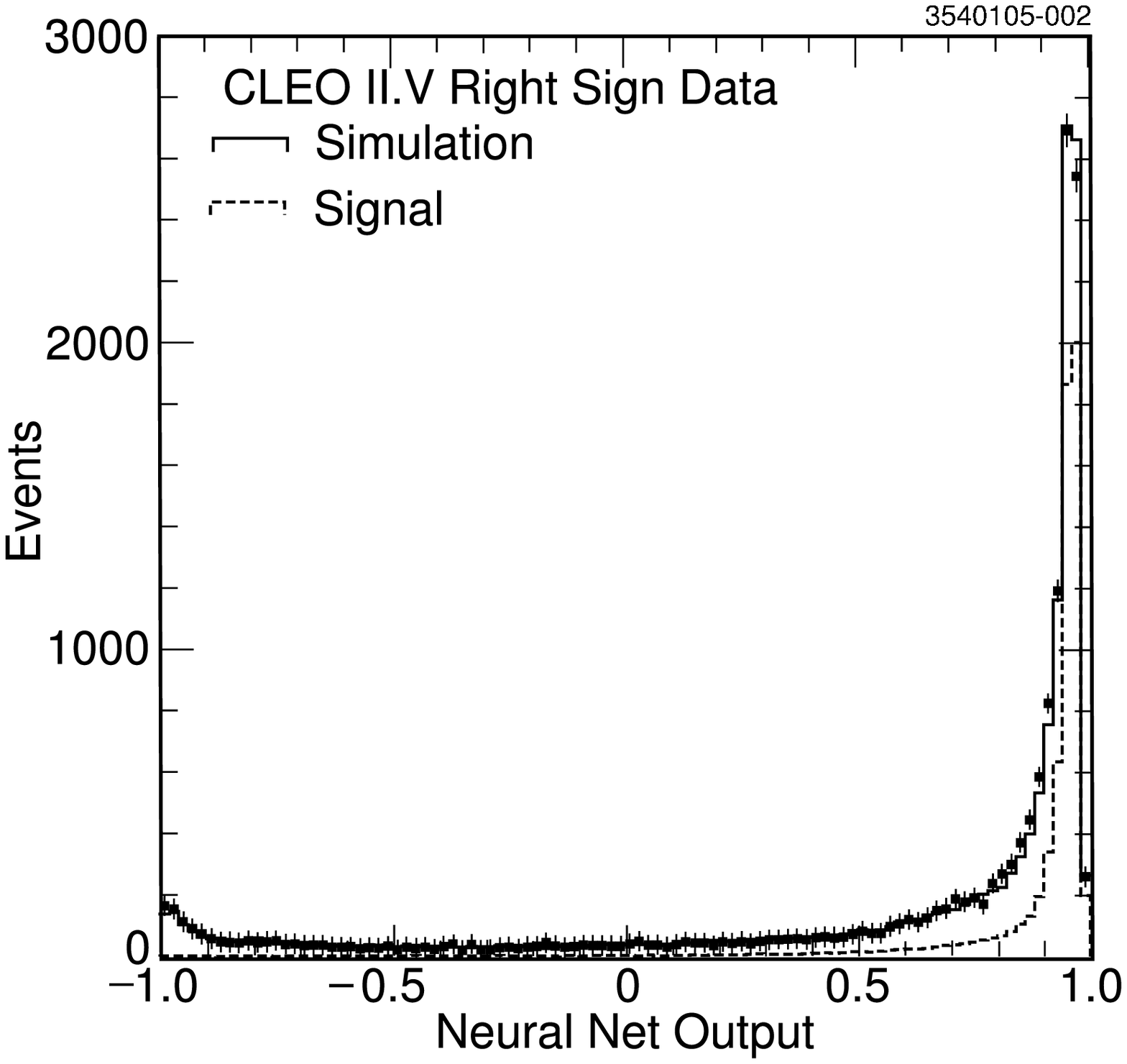}
\includegraphics[height=3.0in]{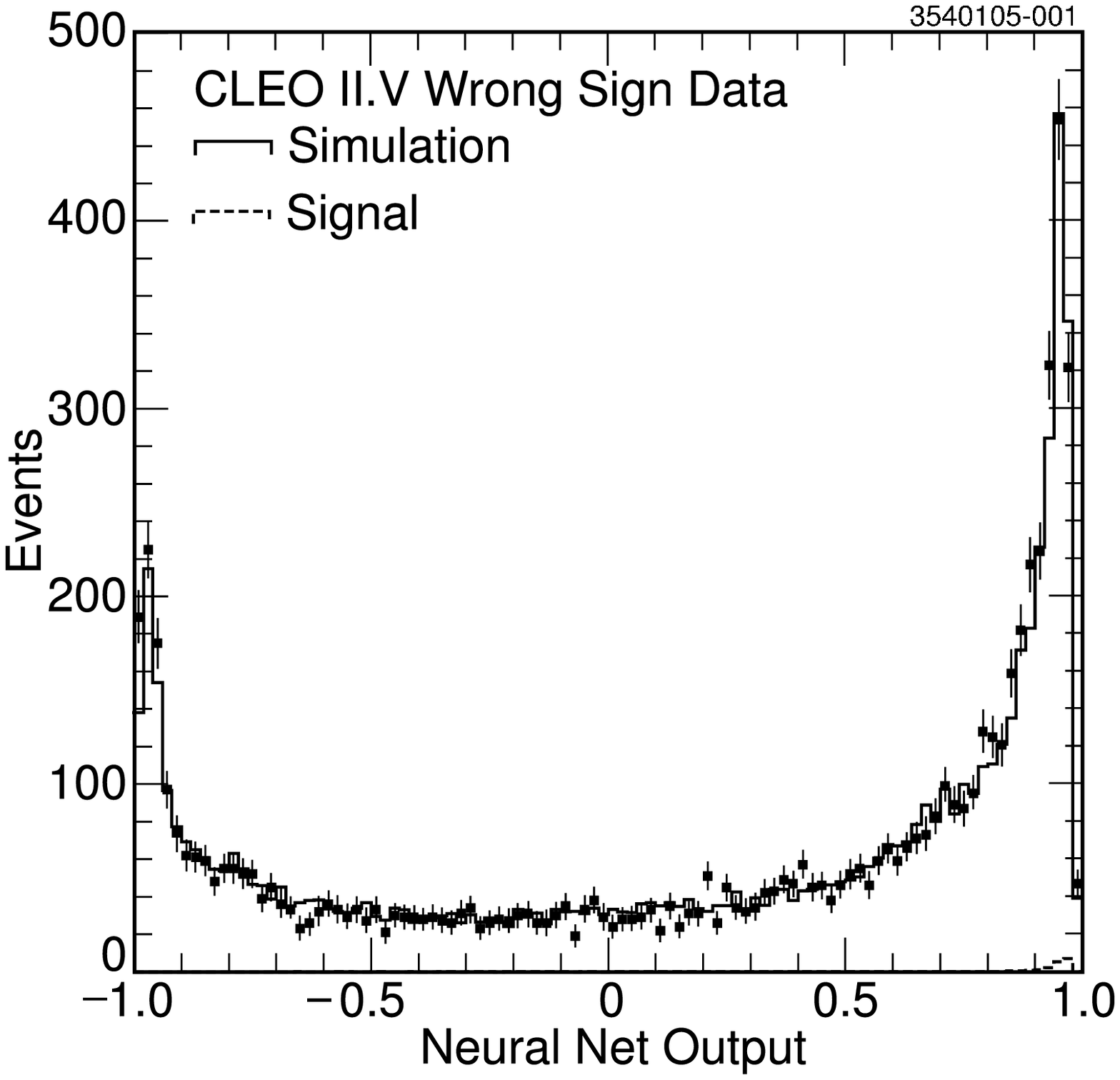}
\caption{The output of the neural net used in the $D^0 \to Ke\nu$ analysis
for the RS (left) and WS (right) samples comparing data (squares) with simulation (full histogram).
Also shown with a dashed histogram is the predicted contribution of the signal.
The very small WS signal is at the central value found by this analysis.}
\label{fig:NN}
\end{figure}

	The $D^0 \to K e \nu$ analysis uses a neural net
to distinguish signal events from background.  All combinations
of electrons, kaons identified loosely via specific
ionization, and $\pi_{\rm soft}$ are considered.  Eighteen variables
are inputs to the net.  They are selected such that our simulation
describes their distribution well, and in general they describe the
kinematics of the $\pi_{\rm soft}$, kaon, and electron candidates
separating random combinations from those produced in the desired
decay chain.
For example one input is the cosine of the angle between the electron and the kaon which
peaks near one for the signal and at both one and negative one for
background.  The net is trained on our simulation and produces an
output near one for signal-like combinations and negative one for background-like
combinations.  Figure~\ref{fig:NN} 
shows the output of the neural net on
RS and WS candidates.  According to the simulation the RS sample is
roughly 42\% signal while we expect the WS to be dominated by
background.
We compare the output of the neural net thoroughly with
the prediction of the simulation by varying both its input and structure.  The
simulation is found to predict very well changes observed in the data.
Combinations with a neural net output of greater than~0.95
are selected for further analysis. 

\begin{figure}
\includegraphics[height=2.9in]{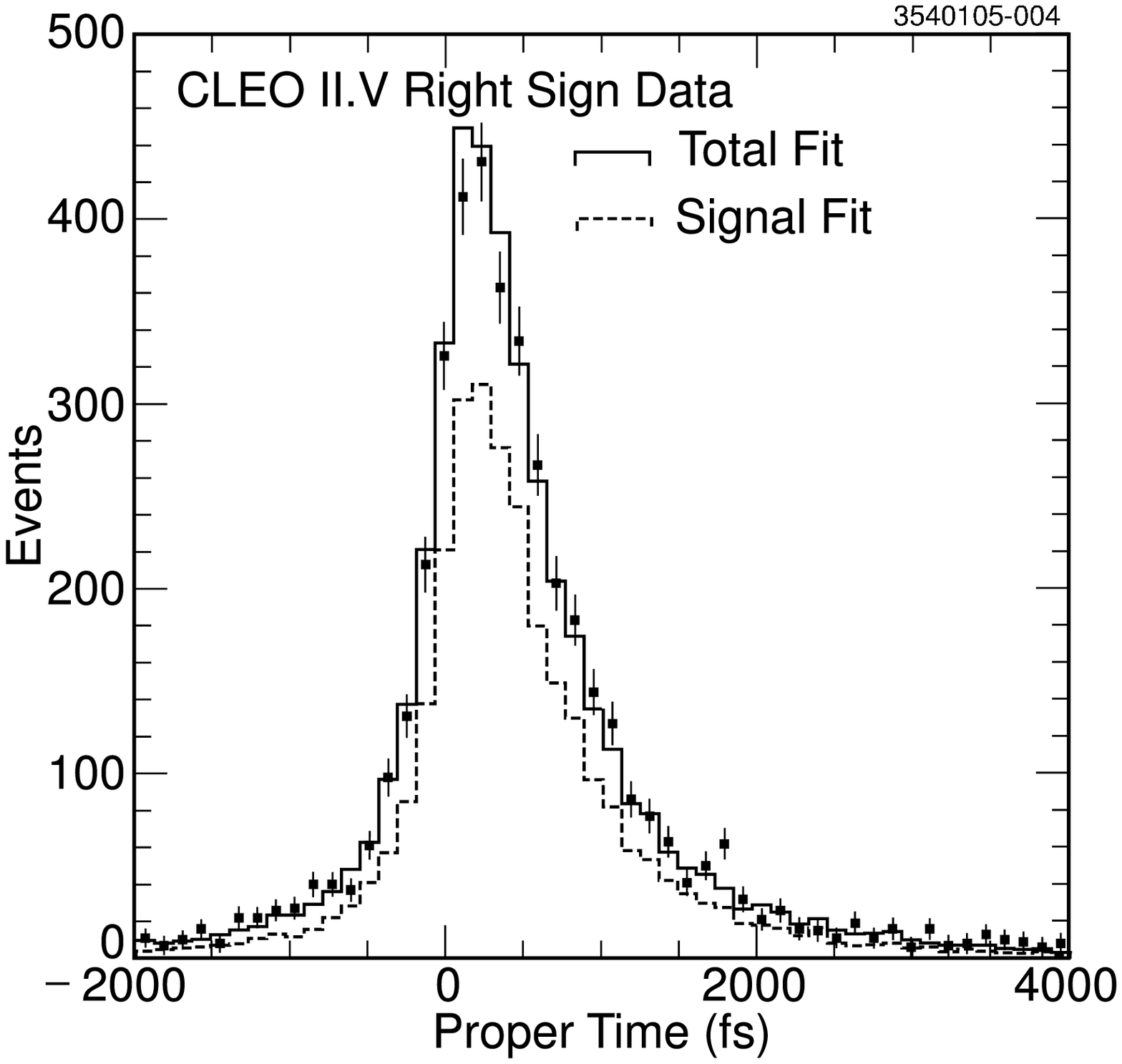}
\includegraphics[height=2.9in]{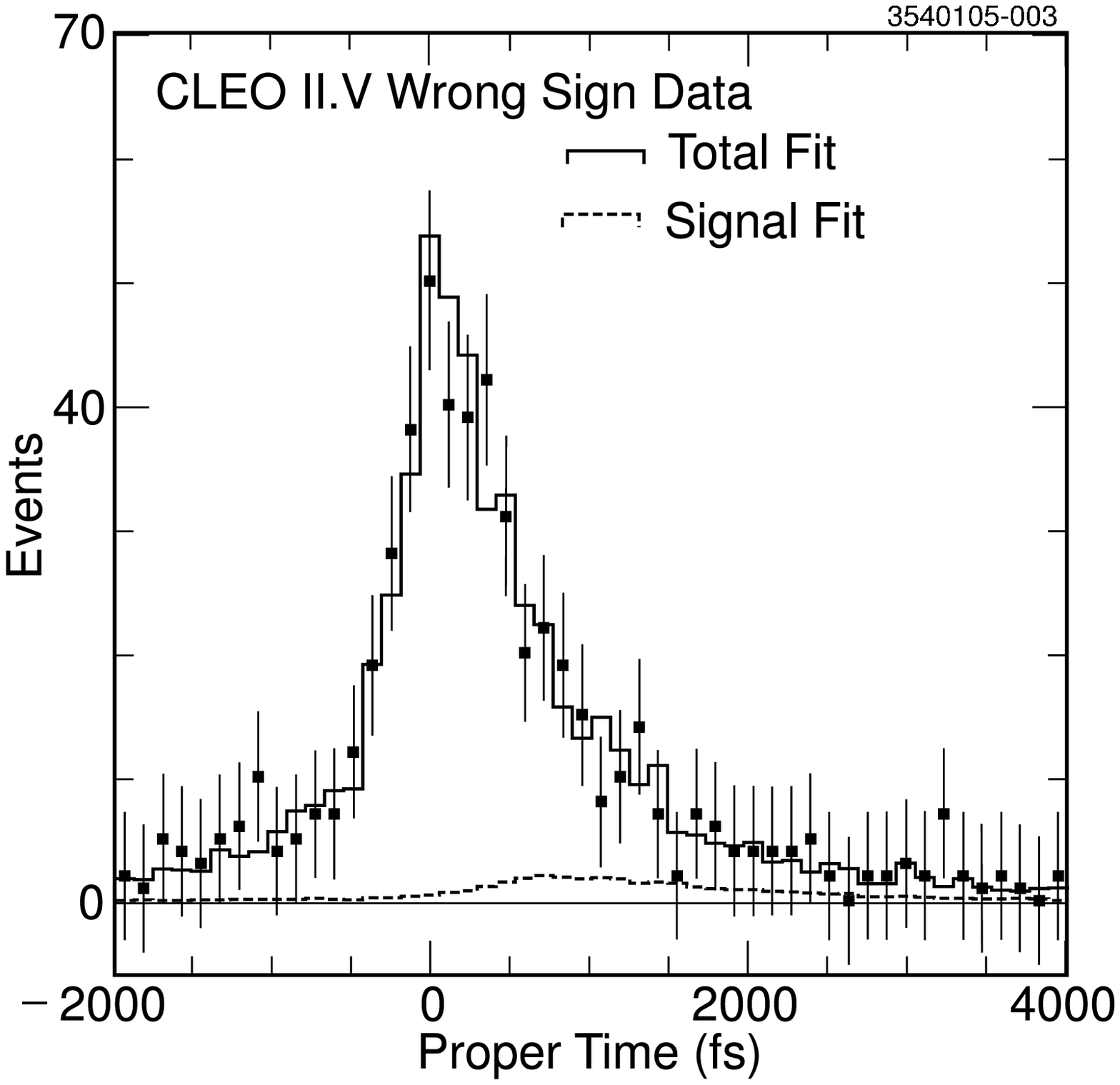}
\caption{The decay time for the $D^0 \to Ke\nu$ analysis
for the RS (left) and WS (right) samples.  The squares show the data, the full histogram shows the fits,
and the dashed histogram the signal contribution to the fits.}
\label{fig:D0life}
\end{figure}

	To measure the decay time of the $D^0$ we refit the tracks
requiring that the kaon and electron come from a common $D^0$ decay vertex,
use the thrust axis of the event as the direction of the $D^0$, and
require that the $D^0$ and $\pi^+_{\rm slow}$ or $\overline{D^0}$ and $\pi^-_{\rm slow}$,
come from a common vertex
constrained to be in luminous region.  This procedure improves the resolution
on the decay time by 30\% to about half a $D^0$ lifetime.
Figure~\ref{fig:D0life} shows the distribution of the decay times for
the RS and WS samples.  Overlaid are fits to a signal plus backgrounds.
Background shapes are taken from the simulation and are dominated by non-charm
events with a significant fraction of non-signal charm decays.  The RS
distribution is dominated by signal, while the WS is consistent with
no signal and thus is dominated by background.  Signal shapes are
also taken from the simulation.  The agreement between data and simulation
is good and the normalizations for the RS signal and the background
in both the RS and WS distributions agree with the predictions of
the simulation.  

	These fits find 2840$\pm$300 RS signal and 31$\pm$21 WS signal
combinations.  In the $D^0 \to K e \nu$  channel we measure $R_M = 0.0110\pm 0.0076$
and see no evidence for mixing.  Systematic uncertainties are discussed below.

	The $D^0 \to K^\ast e \nu$ analysis uses a more traditional approach.
The charged $K^\ast$ is reconstructed in the $K_S^0 \pi$ channel followed by the $K_S^0$ decay
to two charged pions.  These two pions are fit to a common vertex and the mass of
the $\pi\pi$ must be within 16 MeV/c${}^2$ of the expected $K_S^0$ mass. 
Similarly for the $K^\ast$, the $K_S^0$-$\pi$ combination must be within
60 MeV/c${}^2$ of the known mass.  When the $K^\ast$ is combined with
an electron, and the pair is consistent with coming from a $D^0$ decay, the
sample is fairly clean.  The direction of the $D^0$ is determined
with a weighted average of the thrust, the $\pi_{\rm soft}$,
and $K^\ast$-electron combination directions.  This gives a better estimate
of the energy release, $Q$, in the $D^\ast$ decay and signal is
distinguished from background in a fit to the two dimensional
$Q$ versus decay time distribution.  The decay time is improved by a refit similar
to the one described above.

\begin{figure}
\includegraphics{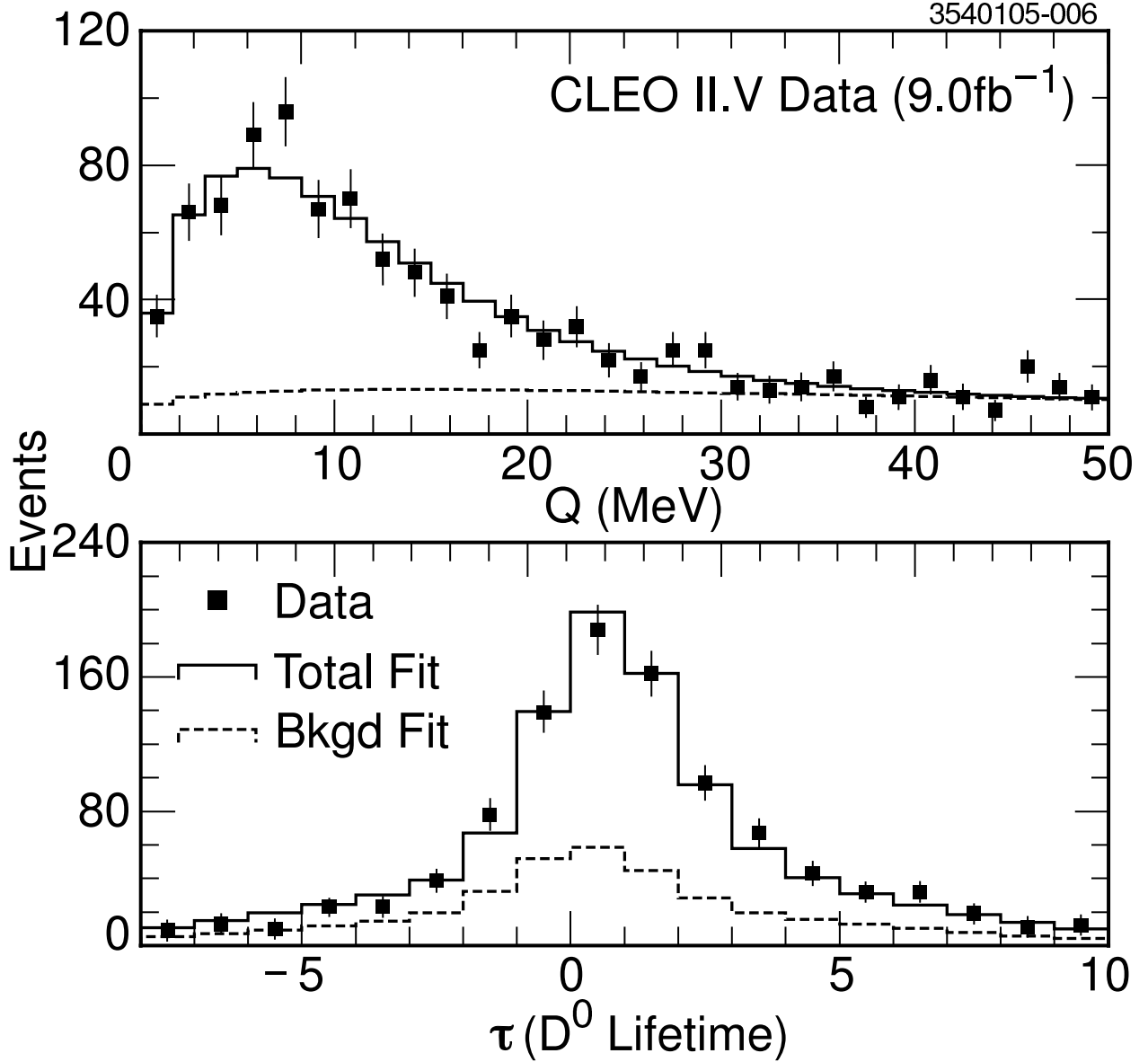}
\caption{The $Q$ (top) and decay time (bottom) distributions
for the $D^0 \to K^\ast e\nu$ analysis
for the RS sample.  The squares show the data, the histograms show projections
of the two dimensional fit,
and the dashed histogram shows the background contribution to the fit.}
\label{fig:2DRS}
\end{figure}

\begin{figure}
\includegraphics{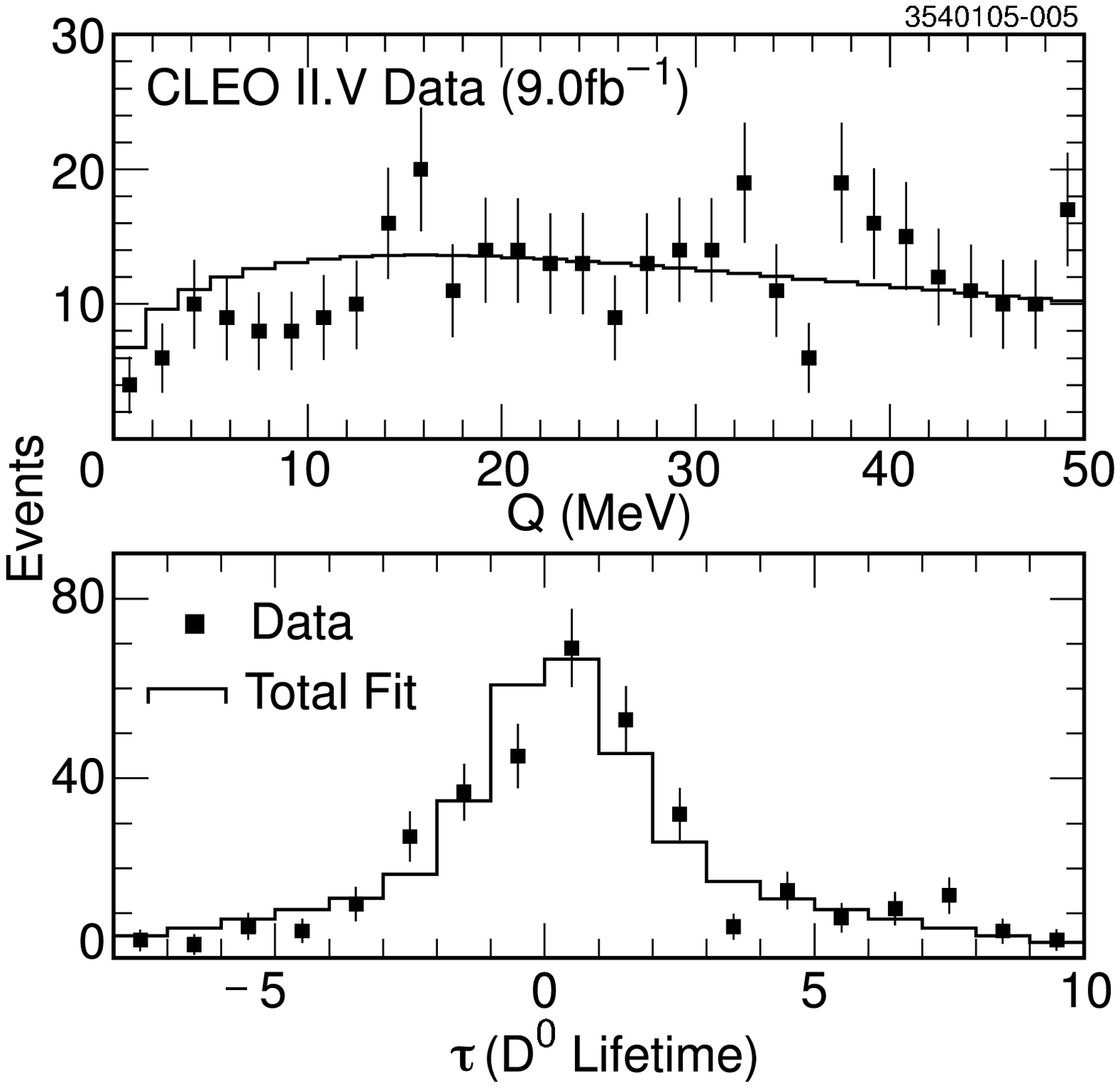}
\caption{The $Q$ (top) and decay time (bottom) distributions
for the $D^0 \to K^\ast e\nu$ analysis
for the WS sample.  The squares shows the data and the histograms shows projections
of the two dimensional fit.}
\label{fig:2DWS}
\end{figure}

	Figures~\ref{fig:2DRS} and~\ref{fig:2DWS} show the projections
of the two dimensional fits on the $Q$ and decay time axes for RS
and WS candidates respectively.  Shapes for signal and background are taken from
the simulation.  Agreement between the data and simulation is good
both in the signal dominated RS sample and the background dominated
WS sample.  The prediction of the simulation is checked using
$D^\ast \to \pi_{\rm slow} D^0 \to K_S^0 \pi\pi\pi^0$ decays found in the data.  The $\pi^0$
is ignored and the same methods
are applied as in the $K^\ast e \nu$ analysis to reconstruct a
two dimensional $Q$ and decay time distribution.  The simulation and the data
agree very well in this check sample.  The simulation also accurately
predicts the size of the observed signal and background in both the RS and
WS samples.

	These fits find 638$\pm$51 RS signal and $-$30$\pm$8 WS signal combinations.
When constrained to find at least zero WS signal the fit returns 0$\pm$2 WS signal
combinations.  The two results yield similar upper limits on $R_M$,
and the latter is used to combine with the $Ke\nu$ channel as it
yields a slightly more conservative upper limit.
In the $D^0 \to K^\ast e \nu$  channel we measure $R_M = 0.0000\pm 0.0031$
and see no evidence for mixing.  Systematic uncertainties are discussed below.

	The two analyses are statistically independent and
consistent; therefore they can be combined.  We combine their central values
weighting by their statistical uncertainties to find $R_M = 0.0016 \pm 0.0029$.

	The two analyses share some common systematic uncertainties.  They
both use the same simulation to model backgrounds and signal shapes.
We take the smaller of the two systematic uncertainties as the uncertainty
on the combined value.
The statistical uncertainty
on simulated signal and background shapes causes an uncertainty of
$\pm 0.0023$ on $R_M$ in the $K^\ast e\nu$ analysis
and $\pm 0.0028$ in the $Ke\nu$ analysis.
The shape of the background in the decay time is parametrized from
the simulation in a similar way in both analyses.  Variations in
this parameterization affect the number of RS and WS signal combinations.
This variation causes an uncertainty of $\pm 0.0014$ on $R_M$ in the $Ke\nu$ analysis
and $\pm 0.0018$ in the $K^\ast e \nu$ analysis.  
The lower of the
two uncertainties, the first in each case, is taken as the uncertainty on
the combined result.

	For systematic uncertainties not shared by the two analyses we add them
by weighting their contribution to the central value.  Specifically uncertainties
in the $Ke\nu$ analysis contribute 15\% of their size to the combined analysis,
while uncertainties in the $K^\ast e\nu$ analysis contribute 85\% of their
size.  
In the $K^\ast e\nu$ analysis
variations of the parametrization of the $Q$ shape for signal and background
add $\pm0.0008$.  
In the $Ke\nu$ analysis the largest contribution to the uncertainty
comes from variations in the electron identification and add $\pm0.0007$ to the combined result.
Details of the lifetime fit (binning, fit range, and $D^0$ direction choice)
add $\pm0.0004$.  Variations in particle identification selections and
details of the refit procedure add $\pm0.0004$.  
Other systematic effects are studied and found to be negligible.

	All the systematic uncertainties are combined in quadrature
to yield a total systematic uncertainty of $\pm 0.0029$.  All of the uncertainties
are summarized in Table~\ref{tab:uncert}.
\begin{table}
\caption{Summary of uncertainties on $R_M$.}
\begin{tabular}{|l|lr|} \hline
Source                        & Size & \\ \hline \hline
Statistics                    & $\pm 0.0029$ & \\ \hline \hline
Simulation Statistics         & & $\pm 0.0023$ \\ \hline
Decay Time Shape              & & $\pm 0.0014$ \\ \hline
$Q$ Parameterization          & & $\pm 0.0008$ \\ \hline
Electron Identification       & & $\pm 0.0007$ \\ \hline
Decay Time Fit Details        & & $\pm 0.0004$ \\ \hline
Particle ID and Refit Details & & $\pm 0.0004$ \\ \hline \hline
Systematic Total              & $\pm 0.0029$ & \\ \hline
\end{tabular}
\label{tab:uncert}
\end{table}

	We see no sign of $D$ mixing in either channel and set limits
on $R_M$ based on our central value of $0.0016\pm0.0029\pm0.0029$ where
the statistical and systematic uncertainties are combined quadratically.
We assume the uncertainty follows a
Gaussian distribution and exclude the unphysical region, $R_M < 0$.
This gives $R_M < 0.0078$ at 90\% C.L. and $R_M < 0.0091$ at 95\% C.L.
We agree with previous measurements and set limits as expected
given the size of our data sample.

We gratefully acknowledge the effort of the CESR staff 
in providing us with excellent luminosity and running conditions.
This work was supported by the National Science Foundation
and the U.S. Department of Energy.

\end{document}